\documentclass[11pt]{article}
\usepackage{array}
\usepackage{amsmath}
\usepackage{amssymb}
\usepackage{graphicx}
\usepackage{amsthm}
\usepackage{latexsym}
\newtheorem{theorem}{Theorem}
\newtheorem{cor}{Corollary}
\newtheorem{definition}{Definition}
\newtheorem{prop}{Proposition}

\newtheorem{example}{Example}

\newcommand{\beq}{\begin{equation}}
\newcommand{\eeq}{\end{equation}}
\newcommand{\barr}{\left[\begin{array}}
\newcommand{\earr}{\end{array}\right]}

\newcommand{\bpf}{\emph{Proof}\/:}
\newcommand{\epf}{\hfill$\Box$}

\newcommand{\bzero}{\ensuremath{B_{0}}}

\newcommand{\bi}{\begin{itemize}}
\newcommand{\ei}{\end{itemize}}
\newcommand{\bnum}{\begin{enumerate}}
\newcommand{\enum}{\end{enumerate}}
\newcommand{\bc}{\begin{center}}

\newcommand{\equ}{\Leftrightarrow}
\begin{document}
\title{Generalization of Boole-Shannon expansion, consistency of Boolean equations and elimination by orthonormal expansion}
\author{Virendra Sule\\Department\ of Electrical Engineering\\Indian Institute of Technology Bombay, India\\vrs@ee.iitb.ac.in}
\maketitle

\begin{abstract}
The well known Boole -Shannon expansion of Boolean functions in several variables (with coefficients in a Boolean algebra $B$) is also known in more general form in terms of expansion in a set $\Phi$ of orthonormal functions. However, unlike the one variable step of this expansion an analogous elimination theorem and consistency is not well known. This article proves such an elimination theorem for a special class of Boolean functions denoted $B(\Phi)$. When the orthonormal set $\Phi$ is of polynomial size in number $n$ of variables, the consistency of a Boolean equation $f=0$ can be determined in polynomial number of $B$-operations. A characterization of $B(\Phi)$ is also shown and an elimination based procedure for computing consistency of Boolean equations is proposed. 
\end{abstract}

Comments: 16 pages, Revised December 4, 2013

Category: cs.CC, cs.SC, ms.RA

ACM class: I.1.2, F.2.2, G.2

MSC class: 03G05, 06E30, 94C10.

\section{Introduction}
For a finite Boolean algebra $B$\footnote{with Boolean operations of addition $+$, multiplication $.$ and complement $'$ satisfying axioms of a Boolean algebra} a Boolean function $f:B^{n}\rightarrow B$ is defined by values of formal Boolean expressions in $n$-variables $\{x_{1},\ldots,x_{n}\}$ using the Boolean operations. For such functions\footnote{hereafter by "function" we shall mean "Boolean function"} $f(x_{1},\ldots,x_{n})$ the following well known representations exist, (known more often as Shannon expansions) \cite{brow}
\beq\label{Bool-Shannon}
\begin{array}{rcl}
f(x_{1},\ldots,x_{n}) & = & x_{i}f(x_{i}=1)+x_{i}'f(x_{i}=0)\\
f(x_{1},\ldots,x_{n}) & = & [x_{i}+f(x_{i}=0)][x_{1}'+f(x_{i}=1)]
\end{array}
\eeq
However, this representation is also believed to have been known to Boole \cite{bool}, hence we shall prefer to call these Boole-Shannon expansions. Thus a function of one variable $f:B\rightarrow B$ has representation $f(x)=ax+bx'$ where $a=f(1)$ and $b=f(0)$. A Boolean equation $f(x)=0$ is said to be \emph{consistent} if there is an element $\alpha$ in $B$ such that $f(\alpha)=0$ in $B$. A well known (and easily provable) result \cite{rude, brow} is that $f(x)=0$ is consistent iff $ab=0$ in $B$. This condition gives a new equation in the \emph{eliminant} $ab$ which does not involve $x$ hence this is an elimination of $x$ from the equation $f=0$. If now $f(X,y)$ is a Boolean function $B^{n+1}\rightarrow B$ and $B(n)$ is the Boolean algebra of functions of $n$ variables $X$ then we have the

\begin{theorem}[Elimination theorem]{\em The equation $f(X,y)=0$ is consistent iff
\[
f_{1}(X)\stackrel{\triangle}{=}f(y=1)f(y=0)=0
\]   
is consistent.}
\end{theorem}

Well known statements of this theorem can be referred from \cite[section 7.3.1]{brow} and \cite[section 2.4]{rude}. By successively eliminating variables from the equation $f=0$ when $f$ (denoted $f_{0}$) is a function of $n$-variables, for each $i=0,1,\ldots$ getting eliminants $f_{i}$ in lesser number of variables the final eliminant (a constant in $B$) gives the consistency condition $f_{m}=0$ for consistency of $f=0$.

\subsection{A generalization of Boole-Shannon expansion}
Let $B(n)$ denote the Boolean algebra of Boolean functions $f:B^{n}\rightarrow B$ in $n$ variables (such a function is also denoted as $f(X)$ where $X$ denotes the $n$ variables \footnote{if $Z$ is a fixed element of $B^{n}$ then $f(Z)$ denotes the value in $B$ of the function $f(X)$ at $Z$}). A set of $m$ functions $\Phi=\{\phi_{1},\ldots,\phi_{m}\}$ in $B(n)$ is said to be \emph{orthogonal} (OG) if $\phi_{i}\phi_{j}=0$ for $i\neq j$ and \emph{orthonormal} (ON) of \emph{order} $m$ if in addition they satisfy
\[
\sum_{i=1}^{m}\phi_{i}(X)=1
\]
In general, over a Boolean algebra $B$, an $n$-tuple $\{x_{1},\ldots,x_{n}\}$ is called an OG system if $x_{i}x_{j}=0$ for $i\neq j$. An OG system is called ON with $n$ its order if $\sum_{i}x_{i}=1$. A generalization of the Boole-Shannon expansion is given by expansion of a function $f$ in $B(n)$ with respect to an ON set $\Phi$ by the well known result \cite[Proposition 3.14.1]{brow}. (The following statement is an existential version).

\begin{prop}{\em
Let $f(X)$ be a Boolean function and $\Phi=\{\phi_{i},i=1\ldots m\}$ be ON in $B(n)$. Then there exist functions $\alpha_{i}$ in $B(n)$ such that $f$ has representation
\beq\label{ONexpansion}
f(X)=\sum_{i}\alpha_{i}(X)\phi_{i}(X)
\eeq
moreover functions $\alpha_{i}(X)\in [f(X)\phi_{i}(X),f(X)+\phi_{i}(X)']$.
}
\end{prop}

(By an interval $[a,b]$ in a Boolean algebra $B$ is meant the set $\{x\in B|a\leq x\mbox{ and }x\leq b\}$ which is not empty only when $a\leq b$\footnote{$a\leq b$ is equivalent to $ab'=0$} in $B$). For instance $\alpha_{i}(X)=f(X)\phi_{i}(X)$ satisfy the expansion but these coefficients are seldom of much use. The first Boole-Shannon expansion follows from above for $\Phi=\{x,x'\}$ as there are unique constants $\alpha_{1}=f(1)$, $\alpha_{2}=f(0)$.

\subsection{Questions addressed}
This article is aimed at addressing the following question arising from the above generalization of the Boole-Shannon representation formula.
\begin{quote}
\item What would be the consistency condition for $f(X)=0$ in terms of the expansion co-efficient functions $\alpha_{i}(X)$ with respect to an ON set of functions $\Phi$?
\end{quote} 

If the equation $f(X)=0$ is consistent then there exists an $n$ tuple $Z$ in $B^{n}$ such that
\[
\begin{array}{rrcl}
 & f(Z) & = & 0\\
\equ & \sum_{i}\alpha_{i}(Z)\phi_{i}(Z) & = & 0\\
\equ & \alpha_{i}(Z)\phi_{i}(Z) & = & 0\forall i=1,\ldots,m\\
\equ & \alpha_{i}(Z) & \leq & \phi_{i}(Z)'\forall i=1,\ldots,m\\
\equ & \prod_{i=1}^{m}\alpha_{i}(Z) & \leq & \prod_{i=1}^{m}\phi_{i}(Z)'=0
\end{array}
\] 
since $\Phi$ is an ON set. This implies that the equation
\[
\prod_{i}\alpha_{i}(X)=0
\]
is consistent. Hence we have a necessary condition. 

\begin{prop}{\em
If $f(X)=0$ is consistent then
\beq\label{NecessaryCond}
\prod_{i=1}^{m}\alpha_{i}(X)=0
\eeq
is consistent.
}
\end{prop}

We thus ask questions such as
\bnum
\item Is this condition sufficient?
\item For what special ON sets $\Phi$ is this sufficient?
\item For what special functions $f$ is this sufficient?
\item What are the analogous expansions and conditions for special problems such as satisfiability of conjunctive (disjunctive) normal forms CNFs (DNFs)?
\enum

We shall address the above questions in this paper. For all notations, background results on Boolean equations and definitions we shall refer to \cite{rude,brow}.

\subsection{Motivation for studying Boolean equation solvers}
Boolean equations arise in a large number of applications in Sciences, Engineering and Operational research. In computer science and electrical engineering problems of verification of software and hardware, artificial intelligence, constraints satisfaction as well as graphs have been traditionally formulated as satisfiability (SAT) problems of CNFs and DNFs. These are special cases of Boolean equation problems. Efficient $0,1$ assignment based solution algorithm such as DPLL \cite{crah,hand} which take advantage of the formulations over the $2$-element Boolean algebra $B_{0}=\{0,1\}$ have been well known for these problems. On the other hand elimination based approaches proposed in \cite{rude} have often been found ignored in the current methods of satisfiability solvers (perhaps for reasons of sequential computational difficulties as surveyed in \cite{crah}). Elimination of a variable applied to a CNF set is equivalent to resolution of its clauses at the pivot variable. Resolution may not be the best method for arbitrary CNF-SAT problems without some prior processing. However elimination based methods are not just applicable for problems without CNF or DNF representation but are also applicable over general Boolean algebras and equations. The potential for development of scalable computational algorithms by elimination based methods for solving very large sized problems by parallel computation is not fully explored. This certainly makes these methods attractive for further exploration. In computational mathematics Boolean equations are also a natural way to formulate problems of rational solutions of polynomial equations over finite fields and problems of solving equations over groups. An enormous recent work is published on algebraic elimination methods such as XL and Grobner basis approaches. Reference \cite{bard} gives an idea of this direction particularly in cryptological research. However, the Boolean approach to equation solving is mathematically very different than algebraic elimination such as in XL or Grobner basis algorithms both of which are extensions of the Gaussian elimination. This is because in Gaussian elimination multiplicative invertibility is the fundamental requirement for solving an equation, while in Boolean algebra there is no concept of a multiplicative inverse. Hence theory of Boolean and Lattice equations proposed in \cite{rud2} forms a different mathematical stream as compared to the algebraic. Hence it is important to explore methods of solving Boolean equations distinctively than algebraic (ring or field theoretic) methods from both theoretical as well as computational point of view.

\section{Orthonormal sets and special Boolean equations}
In this section we gather preliminary results on solution of special Boolean systems which are required for proving further results. These systems are defined by ON sets of functions. Let $B$ be the Boolean algebra of coefficients over which we consider functions of $n$ variables, the Boolean algebra $B(n)$. In $B(n)$ there is the $2^n$ order system of ON functions, that of minterms in $n$ variables denoted
\[
\mu_{j}=x_{1}^{j_{1}}\ldots x_{n}^{j_{n}}
\]
for $j=0,1,\ldots,2^{n}-1$ where $j_{k}=0,1$ and $x^{1}=x$, $x^{0}=x'$. In general an ON set of terms $T=\{t_{j},j=1,\ldots,m\}$ of order $m$ is an ON set which consists of terms of the form
\[
t_{j}=x_{1}^{j_{1}}\ldots x_{n}^{j_{n}}
\]
where $j_{k}=-1,0,1$ and $x^{-1}=1$. If
\beq\label{partition}
\bigcup_{i=1}^{m}M_{i}=\{0,1,\ldots,2^{n}-1\}
\eeq
is a disjoint partition of the $2^n$ indices of the minterms, then $\Phi=\{\phi_{k},k=1,\ldots,m\}$ where
\beq\label{mintermexpansion}
\phi_{k}=\sum_{j\in M_{k}}\mu_{j}
\eeq
is an ON set of order $m$. Conversely, since every function in $B(n)$ has a minterm canonical form, it follows that every ON set can be expressed this way. These developments on ON systems are referred from \cite[chapter 4]{rude}.

\subsection{ON solutions and an associated linear equation}
First, consider the special problem of consistency of a Boolean equation $f(X)=0$ where $f$ is in $B(n)$ for existence of ON solutions for $n\geq 2$. We thus want to determine existence and find $Z$ an ON system of order $n$ in $B$ such that $f(Z)=0$. Denote by $\Delta_{i}$ the ON system $\{0,\ldots,1,\ldots,0\}$ where there is $1$ in the $i$-th place while all other elements are $0$. From the minterm canonical form of $f$ it can be shown that any such ON system $Z=\{z_{1},\ldots,z_{n}\}$ of order $n$ to be a solution of $f(X)=0$ it must satisfy the associated linear equation
\beq\label{associatedlineqn}
\sum_{i=1}^{n}a_{i}\chi_{i}=0
\eeq
where
\[
a_{i}=f(\Delta_{i})
\]
This special equation has following well known condition \cite[theorem 4.7]{rude} for consistency and ON solutions.

\begin{prop}\label{Prop:linsystem}{\em
The linear equation (\ref{associatedlineqn}) has an ON solution $\{z_{1},\dots,z_{n}\}$ iff
\beq\label{condforONeqn}
\prod_{1=1}^{n}a_{i}=0
\eeq
one solution is $z_{1}=a_{1}'$,
\[
z_{i}=a_{1}a_{2}\ldots a_{i-1}a_{i}'
\]
for $i=2,\ldots,n$.
}
\end{prop}

Note that there are several other ON solutions such as
\[
\begin{array}{rcl}
z_{\sigma(1)} & = & a_{\sigma(1)}'\\
z_{\sigma(i)} & = & a_{\sigma(1)}a_{\sigma(2)}\ldots a_{\sigma(i-1)}a_{\sigma(i)}'
\end{array}
\]
as $\sigma$ varies over all permutations in $S_{n}$. It will shown elsewhere that these are all solutions of the equation (\ref{associatedlineqn}).

\subsubsection{Co-orthonormal systems, solutions and special equations}
As a dual of above developments consider the consistency of the equation $f(X)=1$ for solutions which are called \emph{co-orthonormal} (co-ON). We call a system $\beta_{i},i=1,\ldots,m$ in $B$ to be co-ON of order $m$ if $\beta_{i}+\beta_{j}=1$ for $i\neq j$ and
\[
\prod_{i}\beta_{i}=0
\]
If $\beta_{i}$ is an ON system then $\beta{i}'$ is a co-ON system. By dual arguments it can be shown that if the equation $f(X)=1$ has an co-ON solution $Y=(y_{1},\ldots,y_{n})$ then $Y$ satisfies the associated dual-linear equation
\beq\label{associatedcoONeqn}
\prod_{i}(b_{i}+\xi_{i})=1
\eeq

The consistency of the dual-linear equation (\ref{associatedcoONeqn}) is given by the dual result to the above proposition. We omit the proof as it can be easily developed from the previous case by dual arguments.
  
\begin{prop}{\em
The equation
\[
\prod_{i}(b_{i}+\xi_{i})=1
\]
has a co-ON solution iff
\[
\sum_{i=1}^{n}b_{i}=1
\]
One solution is 
\[
\begin{array}{rcl}
\xi_{1} & = & b_{1}'\\
\xi_{i} & = & b_{1}+b_{2}+\ldots+b_{i-1}+b_{i}'
\end{array}
\]
for $i=2,\ldots,n$. 
}
\end{prop} 

\subsection{Single equation in minterms}
Consider next a special equation involving minterms. Let $\{\mu_{i},i=0.\ldots,2^{n}-1\}$ be the set of all minterms in $n$ variables $X$ and $B$ be a Boolean algebra. When is the equation
\beq\label{ONmintermeqn}
\sum_{i=0}^{2^{n}-1}\alpha_{i}\mu_{i}(X)=0
\eeq
consistent, where $\alpha_{i}$ are constants in $B$? The consistency condition and a solution of \cite[theorem 4.7]{rude} as used above in proposition \ref{Prop:linsystem} also leads to the following proposition\footnote{Professor Rudeanu has pointed out that this proposition is the well known Boole-Schroader theorem \cite[theorem 2.3]{rude}. The proof is still included for convenience of ready reference of the reader} 

\begin{prop}{\em
The Boolean equation (\ref{ONmintermeqn}) in minterms $\mu_{i}(X)$ in $n$ variables $X$ is consistent iff
\beq\label{ConofONmintermeqn}
\prod_{i=0}^{2^{n}-1}\alpha_{i}=0
\eeq
}
\end{prop}

\bpf
If this equation is consistent then there exists $Z=\{z_{1},\ldots,z_{n}\}$ in $B^n$ and constants $\beta_{i}$ such that
\[
\mu_{i}(Z)=\beta_{i}
\]
and $\beta_{i}$ satisfy
\[
\sum_{i=0}^{2^{n}-1}\alpha_{i}\beta_{i}=0
\]
Also $\beta_{i}$ is an ON system of order $2^n$. The equation (\ref{ONmintermeqn}) when minterms are evaluated at values $X=Z$ then implies that
\[
\alpha_{i}\leq\beta_{i}'\mbox{ for }i=0,\ldots,2^{n}-1
\]
hence the necessary condition for consistency of (\ref{ONmintermeqn}) is
\[
\prod_{i=0}^{2^{n}-1}\alpha_{i}=0
\]
Conversely if this condition holds, then define
\[
\begin{array}{rcl}
\beta_{0} & = & \alpha_{0}'\\
\beta_{i} & = & \alpha_{1}\alpha_{2}\ldots\alpha_{i-1}\alpha_{i}'
\end{array}
\]
for $i\geq 1$. Then $\beta_{i}$ is an ON system of order $2^n$ as $\beta_{i}\beta_{j}=0$ for $i<j$ and
\[
\sum_{i}\beta_{i}=\sum_{i}\alpha_{i}'=1
\]
moreover $\beta_{i}$ satisfy
\[
\sum_{i=0}^{2^{n}-1}\alpha_{i}\beta_{i}=0
\]
Consider the system
\beq\label{mintermsys}
\mu_{i}(X)=\beta_{i}
\eeq
in $n$ variables $X$. This has a solution
\[
x_{i}=\sum_{j,\mu_{j}\leq x_{i}}\beta_{j}
\]
due to the expansion formulas for variables $x_{i}$ in terms of minterms. Hence the equation (\ref{ONmintermeqn}) is consistent.
\epf

We use this condition below for studying consistency of general systems defined by ON functions.

\subsection{Boolean system in general ON functions}
Next consider the Boolean system defined by an ON set of functions $\Phi$ of order $m$ in $B(n)$. This is a system of the form
\beq\label{ONeqnsystem}
\phi_{i}(X)=\beta_{i}
\eeq
for constants $\beta_{i}$ in $B$ and $\phi_{i}$ in $\Phi$. This is a system of $m$ equations in $n$ variables. However the unknowns appear as arguments of an ON set of functions. This makes such equations special.

\begin{prop}\label{prop:ONeqnsystem}{\em
The system of equations (\ref{ONeqnsystem}) defined by an ON set $\Phi$ of order $m$ is consistent iff $\{\beta_{1},\ldots,\beta_{m}\}$ is an ON system of order $m$ in $B$.
}
\end{prop}

\bpf
Necessity is obvious. Conversely let $\beta_{i}$ be an ON system of order $m$. Since $\phi_{i}(X)$ are ON functions, there exists a disjoint partition and expansion (\ref{partition}), (\ref{mintermexpansion}) in minterms of functions $\phi_{i}(X)$ such that
\[
\phi_{i}(X)=\sum_{j,j\in M_{i}}\mu_{j}(X)=\beta_{i}
\]
For each $i$ we choose arbitrary but unique $j=k_{i}$ and consider equations
\[
\begin{array}{rcl}
\mu_{k_{i}}(X) & = & \beta_{i}\\
\mu_{j}(X) & = & 0
\end{array}
\]
for all $j\in M_{i}, j\neq k_{i}$ and $i=1,\ldots,m$. This system is however of the form (\ref{mintermsys}) and is consistent if the constants on the right hand side are ON. If the $m$ constants $\beta_{i}$ are ON then the $2^n$ constants
\[
\begin{array}{rcl}
\alpha_{k_{i}} & = & \beta_{i}\\
\alpha_{j} & = & 0
\end{array}
\]
where $k_{i}\neq j\in M_{i}$ and $i$ is the index of partitions, are also ON. Hence there is a solution
\[
x_{i}=\sum_{j,\mu_{j}\leq x_{i}}\alpha_{j}
\]
which is also a solution of the system (\ref{ONeqnsystem}) 
\epf

It is worth looking at an example to illustrate the above solution.

\begin{example}{\em
Consider following ON system in terms of ON functions of three variables $x,y,z$ over a Boolean algebra $B$.
\[
\begin{array}{rclll}
\phi_{1} & = & x'y'z'+xy'z+xyz & = & \beta_{1}\\
\phi_{2} & = & x'y'z+x'yz+xyz' & = & \beta_{2}\\
\phi_{3} & = & x'yz'+xy'z' & = & \beta_{3}
\end{array}
\]
where $\beta_{i}\beta_{j}=0$ for $i\neq j$ and $\beta_{1}+\beta_{2}+\beta_{3}=1$. Define the system
\[
\begin{array}{lll}
x'y'z'=0 & xy'z=\beta_{1} & xyz=0\\
x'y'z=\beta_{2} & x'yz=0 & xyz'=0\\
x'yz'=\beta_{3} & xy'z'=0 & 
\end{array}
\]
This system has the solution
\[
\begin{array}{rclll}
x & = & x(y'z'+y'z+yz'+yz) & = & 0+\beta_{1}+0+0=\beta_{1}\\
y & = & y(x'z'+x'z+xz'+xz) & = & \beta_{3}+0+0+0=\beta_{3}\\
z & = & z(x'y'+x'y+xy'+xy) & = & \beta_{2}+0+\beta_{1}+0=\beta_{1}+\beta_{2}
\end{array}
\]
For these solutions we have, using the ON relations of $\beta_{i}$
\[
\begin{array}{rcl}
\phi_{1}(x,y,z) & = & \beta_{1}'\beta_{3}'\beta_{2}'+\beta_{1}\beta_{3}'(\beta_{1}+\beta_{2})+0\\
 & = & 0+\beta_{1}\beta_{3}'\\
 & = & \beta_{1}(\beta_{1}+\beta_{2})\\
 & = & \beta_{1}
\end{array}
\]
similar computations verify that $\phi_{2}(x,y,z)=\beta_{2}$, $\phi_{3}(x,y,z)=\beta_{3}$.
}
\end{example}   

\section{Consistency in terms of ON expansion for the class $B(\Phi)$}
We now address the central question of this paper. A Boolean function $f(X):B^{n}\rightarrow B$ when expanded in terms of an ON set $\Phi$ of order $m$ as in (\ref{ONexpansion}) is
\[
f(X)=\sum_{i=1}^{m}\alpha_{i}(X)\phi_{i}(X)
\]
We want to determine the consistency condition for the equation
\[
f(X)=0
\]
in terms of the expansion coefficients $\alpha_{i}(X)$ having noted the necessary condition (\ref{NecessaryCond}). We can establish sufficiency of this condition for a restricted class of Boolean functions defined as follows.

\begin{definition}[Class $B(\Phi)$]{\em Let $B$ be a Boolean algebra and $\Phi=\{\phi_{1},\ldots,\phi_{n}\}$ be an ON set of functions in $B(n)$. A Boolean function $f$ in $B(n)$ is said to be of class $B(\Phi)$ if $f(X)$ has an expansion  (\ref{ONexpansion}) in which all $\alpha_{i}$ are constants i.e. elements of $B$}.
\end{definition}

For this class of functions we have

\begin{theorem}\label{th:consistencycond}{\em
Let $f$ be in $B(\Phi)$ with ON expansion
\[
f(X)=\sum_{i=1}^{m}\alpha_{i}\phi_{i}(X)
\]
Then $f(X)=0$ is consistent iff
\[
\prod_{i}\alpha_{i}=0
\]
}
\end{theorem}

\bpf
We need only prove sufficiency of the condition. If the condition holds then the associated equation
\[
\sum_{i=1}^{m}\alpha_{i}\chi_{i}=0
\]
has an ON solution 
\[
\chi_{i}=\beta_{i}
\]
where $\beta_{i}$ is an ON system of order $m$ in $B$. We may thus consider consistency of the ON equation system (\ref{ONeqnsystem})
\[
\phi_{i}(X)=\beta_{i}
\]
at any one such solution. But this is consistent by proposition (\ref{prop:ONeqnsystem}). Hence there exists an $n$-tuple $Z$ in $B^n$ such that
\[
f(Z)=\sum_{i=1}^{m}\alpha_{i}\phi_{i}(Z)=0
\]
which shows that the condition is sufficient. 
\epf

\subsection{Variable decomposition and elimination interpretation}
From a computational point of view the above result can be interpreted for the purpose of decomposition of variables and interpretation for elimination. Let $f$ be a Boolean function in $B(N)$ where the $N$ variables $X$ can be partitioned as $X=\{X_{1},X_{2}\}$ in two subsets of sizes $n_{1}$, $n_{2}$. Let $\Phi(X_{1})$ denote a set of ON functions in $X_{1}$ variables of order $m_{1}$ and $B_{X_{2}}$ denotes the Boolean algebra of functions in $B(n_{2})$ in $X_{2}$ variables. Hence we have

\begin{cor}{\em
If $f$ belongs to $B_{X_{2}}(\Phi(X_{1}))$ and has an expansion
\[
f(X)=\sum_{i=1}^{m_{1}}\alpha_{i}(X_{2})\phi_{i}(X_{1})
\]
then $f(X)=0$ is consistent iff the equation
\[
\prod_{i=1}^{m_{1}}\alpha_{i}(X_{2})=0
\]
is consistent.
}
\end{cor}  

\bpf
If $f(X)=0$ is consistent then there exist $Z_{1}$ in $B^{n_{1}}$ and $Z_{2}$ in $B^{n_{2}}$ such that
\[
\sum_{i=1}^{m_{1}}\alpha_{i}(Z_{2})\phi_{i}(Z_{1})=0
\]
hence from the associated linear equation as above it follows that
\[
\prod_{i=1}^{m_{1}}\alpha_{i}(Z_{2})=0
\]
which shows that the condition is necessary. Conversely this condition is sufficient for consistency of the associated equation
\[
\sum_{i=1}^{m_{1}}\alpha_{i}(Z_{2})\chi_{i}=0
\]
for an ON solution $\chi_{i}=\beta_{i}$. Given that this is an ON system in $B$ the system of equations
\[
\phi_{i}(X_{1})=\beta_{i}
\]
is consistent as shown in proposition (\ref{prop:ONeqnsystem}). Hence there exists $Z_{1}$ such that 
\[
f(Z_{1},Z_{2})=\sum_{i=1}^{m_{1}}\alpha_{i}(Z_{2})\phi_{i}(Z_{1})=0
\]
which shows that $f(X)=0$ is consistent. This proves sufficiency.
\epf

\subsection{Elimination interpretation}
The statement of the last proposition shows that for the class of functions $f(X_{1},X_{2})$ in $B_{X_{2}}(\Phi(X_{1}))$ consistency of $f$ is equivalent to that of the equation
\[
\prod_{i=1}^{m_{1}}\alpha_{i}(X_{2})=0
\]
which has no presence of $X_{1}$ variables. This statement is thus a generalization of the elimination theorem which originally follows from Boole-Shannon expansion for this restricted class of functions. We shall call the product in the left hand side of the above equation an \emph{eliminant} of $f(X)$ after elimination of variables $X_{1}$.

\subsection{Characterization of $B(\Phi)$}
Given a Boolean function $f$ and an ON set $\Phi$ in $B(n)$, the ON expansion (\ref{ONexpansion}) in terms of the set $\Phi$ becomes
\[
f(X)=\sum_{\phi_{i}\in\Phi}\alpha_{i}(X)\phi_{i}(X)
\]
The co-efficient functions $\alpha_{i}(X)$ satisfy the interval 
\beq\label{coeffrange}
\alpha_{i}(Z)\in[f(Z)\phi_{i}(Z),f(Z)+\phi_{i}(Z)']
\eeq 
for all $Z$ in $B^{n}$. Hence the function belongs to the class $B(\Phi)$ iff there exist constants in this interval. This is made precise in the following.

\begin{prop}\label{prop:coeffrange}{\em Let $\Phi$ be an ON set and $f$ be in $B(n)$. Then $f$ is in $B(\Phi)$ iff
\beq\label{minmaxineq}
\sum_{A\in\{0,1\}^{n}}f(A)\phi(A)\leq\prod_{A\in\{0,1\}^{n}}(f(A)+\phi(A)')
\eeq
}
for all $\phi$ in $\Phi$.
\end{prop}

\bpf
Let $a$ be a constant and $\phi$ an element of $\Phi$ such that $a$ lies in the range (\ref{coeffrange}). From the lower inequality
\[
f(Z)\phi(Z)\leq a\forall Z\in B^{n}  
\]
From the well known range formula \cite[Theorem 2.4]{rude} we have
\[
\max_{x\in B^{n}}f(x)\phi(x)=\sum_{A\in\{0,1\}^{n}}f(A)\phi(A)
\]
from which it follows that
\[
\sum_{A\in\{0,1\}^{n}}f(A)\phi(A)\leq a
\]
Similarly from the range formula we have
\[
\min_{x\in B^{n}}(f(x)+\phi(x)')=\prod_{A\in\{0,1\}^{n}}(f(A)+\phi(A)')
\]
Hence a constant $a$ in the range (\ref{coeffrange}) implies from the higher inequality
\[
a\leq\prod_{A\in\{0,1\}^{n}}(f(A)+\phi(A)')
\]
Hence the claimed inequality holds.
\epf

\subsection{Characterization over the Boolean algebra $B_{0}$}
In this section we consider the above problem of characterizing $B(\Phi)$ when $B$ is the two element Boolean algebra $B_{0}=\{0,1,+,.,'\}$ also known as the switching algebra. Consistency of Boolean equations over \bzero\ is decided by $0,1$ assignments of the variables and all functions in $\bzero(n)$ have values $0,1$. For the expansion (\ref{ONexpansion}) of a function $f$ in $\bzero(\Phi)$ the constants can be either $0$ or $1$. Hence one gets a concrete characterization of this class. We recall a definition, if $f(X)$ is a Boolean function in $B(n)$ we say that $f(X)=0$ is a \emph{Tautology} if $f(X)$ is the zero function i.e. $f(X)=0$ has the solution set $B^n$.

\begin{cor}{\em Let $\Phi$ be an ON set in $\bzero(n)$ and $f$ is in $\bzero(\Phi)$ with expansion
\[
f(X)=\sum_{i}a_{i}\phi_{i}
\]
Then
\[
\begin{array}{rcll}
a_{i} & = & 0 & \mbox{ iff }f\phi_{i}=0 \mbox{ is a Tautology}\\
 & = & 1 & \mbox{ iff }f'\phi_{i}=0\mbox{ is a Tautology}
\end{array}
\]
}
\end{cor}     

\bpf
It follows from Proposition \ref{prop:coeffrange} that the coefficients belong to the range
\[
a_{i}\in [\sum_{A\in\{0,1\}^{n}} (f(A)\phi_{i}(A)),\prod_{A\in\{0,1\}^{n}}(f(A)+\phi_{i}(A)')]
\]
Let $a_{i}=0$. Then from lower inequality this implies
\[
f(A)\phi_{i}(A)=0
\]
for all $A$ in $\{0,1\}^{n}$. This is equivalent to $f\phi_{i}=0$ being a Tautology since this is a Boolean function. Conversely if $f\phi_{i}=0$ is a Tautology then $0$ belongs to the range $[f\phi_{i},f'+\phi_{i}]$ hence $a_{i}=0$ is an admissible co-efficient in expansion term of $\phi_{i}$.

On the other hand let $a_{i}=1$. Then from the upper inequality it follows that
\[
f(A)+\phi_{i}(A)'=1
\]
for all $A$ in $\{0,1\}^{n}$. Hence this being a Boolean function, $f+\phi_{i }=1$ is a Tautology. Taking compliment $f'\phi_{i}=0$ is a Tautology. Conversely if this condition holds then $a_{i}=1$ exists in the co-efficient range. Hence $a_{i}=1$ in an admissible co-efficient in expansion term of $\phi_{i}$  
\epf

Next, we also find a consequence of specializing the result of Theorem \ref{th:consistencycond} over the algebra \bzero.

\begin{cor}{\em
Let $\Phi$ be an ON set in $\bzero(n)$ and $f$ be in $\bzero(\Phi)$. Then
\begin{enumerate}
\item $f(X)=0$ is consistent iff $f\phi_{i}=0$ is a Tautology for some $i$.
\item $f(X)=1$ is consistent iff $f'\phi_{i}=0$ is a Tautology for some $i$.
\end{enumerate}
}
\end{cor}

\bpf
First consider the equation $f(X)=0$. By specialization of the consistency condition of Theorem \ref{th:consistencycond} to \bzero\ it follows that this equation is consistent iff $a_{k}=0$ for some $k$ where $a_{k}$ is a co-efficient in the ON expansion (\ref{ONexpansion}). Since
\[
a_{k}\in [f(X)\phi_{k}(X),f(X)+\phi_{k}(X)']
\]
it follows that $f(X)\phi_{k}(X)=0$ for all $X$ in $\bzero^n$. This proves the first statement. The second statement can be proved on analogous lines by considering the equation $f(X)=1$.  
\epf

\section{Expansion relative to ON terms}
Let $T$ denote a set of ON terms. Then an expansion of $f(X)$ in $T$ is given by \cite[theorem 3.15.1]{brow}
\[
f(X)=\sum_{i,t_{i}\in T}(f/t_{i})(X)t_{i}(X)
\]
where $f/t_{i}$ denotes the function $f(X)|_{t_{i}(X)=1}$ (as $t_{i}(X)$ is a term $t_{i}=1$ has a unique solution). Hence when the function $f$ belongs to $B(T)$ the unique constant $\alpha_{i}=f(t_{i}(X)=1)$ in the expansion. For the case of $f$ in $\bzero(T)$ this has the consistency condition

\begin{cor}{\em Let $T$ be an ON set of terms in $\bzero(n)$ and $f$ is in $\bzero(T)$. Then $f(X)=0$ is consistent iff $f(t_{i}=1)=0$ for at least one $i$. Hence $x_{j}^{\beta_{j}}=1$, $\beta_{j}=\{0,1\}$ is a solution for all $j$ such that $t_{i}\leq x_{j}^{\beta_{j}}$.
}
\end{cor}

Thus it follows that construction of an ON set of terms $T$ for a given function $f(X)$ such that $f$ belongs to $\bzero(T)$ is advantageous in deciding consistency of $f(X)=0$ and computing a solution when it is consistent.   

\section{Computational procedure}
We conclude now with a computational procedure for solution of a Boolean equation $f(X)=0$ using successive ON expansions by decomposing the variables $X$ over the Boolean algebra \bzero. First we note following important remarks concerning the elimination of variables which results from ON expansion.
\bnum
\item Given a function $f(X)$ in $B(n)$ there always exists an ON (exponential sized) set $\Phi$ such that $f$ belongs to $B(\Phi)$. This exponential set is that of minterms in $X$. However, a polynomial sized set $\Phi$ such that $f$ belongs to $B(\Phi)$ may also exist. The consistency condition of Theorem \ref{th:consistencycond} shows that if $\Phi$ is of polynomial size in $n$ the number of variables then the consistency checking and computation of a solution to $f(X)=0$ can be achieved in polynomial number of $B$ operations. Hence problems $f(X)=0$ for $f$ in $B(\Phi)$ can be called polynomial time solvable for a polynomial sized $\Phi$.
\item In $m$ variables the smallest size of an ON set of terms is $m+1$. (For instance $\phi_{i}=x_{1}\ldots x_{i-1}x_{i}'$ for $i=1,\dots,m$). By permutation of the variables and their complements there exist $2m!$ ON sets of terms of size $m+1$. Hence in general constructing an ON set of polynomial size for a given function $f$ such that $f$ belongs to $B(\Phi)$ is not feasible by brute force search for $m$ other than very small.
\item The variables $X$ may be split in disjoint union $X=X_{1}\bigcup,\ldots,X_{r}$ where each set $X_{k}$ is of size $m_{k}$. By the first remark there exist ON sets $\Phi_{1},\Phi_{2},\dots,\Phi_{r}$ such that the eliminants of $f=f_{0}$ given by $f_{1},\ldots,f_{r}$ are in reducing variables and can be computed in Boolean algebras $B(n-n_{k})$ where $n_{k}=m_{1}+\ldots+m_{k-1}$. Hence from the second remark above on polynomial number of operations it follows that if $\Phi_{k}$ are of polynomial size in $m_{k}$ then the consistency of $f=0$ can be computed in polynomial time.    
\enum 

We conclude with the following conceptual procedure for elimination of variables as indicated in remark 3 above, by splitting of variables, construction of an ON set and expansion of the function in the set, in the special case of the equation $f(X)=0$ when $f$ belongs to $\bzero(n)$.

\subsection{Procedure for elimination of variables}
Consider $f(X)$ in $\bzero(n)$ with variable decomposition $X=X_{1}\bigcup X_{2}\ldots\bigcup X_{r}$. Let $\Phi_{k}$ denote the ON set of minterms in $X_{k}$. An elimination procedure is

\subsubsection*{Procedure for elimination}\emph{Input}\/: $f(X)$
\bnum
\item $f_{0}=f(X)$, split as disjoint union $X=X_{1}\bigcup,\ldots,X_{r}$.
\item Set $k=1$. Construct ON $\Phi_{k}$ in $X_{k}$ variables, set $Y_{k}\leftarrow X_{k+1}\bigcup\ldots X_{r}$ such that expansion of $f_{k-1}$ in $\Phi_{k}$ is
\[
f_{k-1}=\sum_{\phi\in\Phi_{k}}\alpha_{\phi}(Y_{k})\phi
\]
\item Compute (this is the resolution or elimination step) 
\[
f_{k}=\prod_{\phi\in\Phi_{k}}\alpha_{\phi}(Y_{k})
\]
\item $k\leftarrow k+1$. 
\item Repeat from 2 untill $k=r$.
\item Consistent when $f_{r}=0$ else not consistent.
\enum

\subsection{Computation of a solution}
The elimination of variables of the above procedure leads to a procedure for computation of a solution by back substitution as follows. The step before the last returns the ON expansion
\[
f_{r-1}=\sum_{i}a_{i}\phi_{i}(X_{r})
\]
The equation $f=0$ is consistent iff one of $a_{i}=0$. Hence the ON system 
\[
\begin{array}{rcl}
\phi_{i}(X_{r}) & = & 1\\
\phi_{j}(X_{r}) & = & 0
\end{array}
\]
for $j\neq i$, gives a solution assignment $\xi_{i}$ in $X_{r}$ variables. Back substitution of this assignment in $f_{r-2}$ gives one of the expansion co-efficient zero at this assignment. Then choose this term (say $k$-th) in the expansion where this coeffcient $\alpha_{k}(\xi_{i})$ is zero and again consider the system with $\phi_{k}(X_{r-1})=1$ to get the assignment of $X_{r-1})$ and continue backwards.  We illustrate this procedure with following example. 

\begin{example}{\em $f(x,y,z)=x+xy'z+xy+x'z'+xy'+xyz$. Let $\Phi_{1}(y,z)=\{yz,y'z,yz',y'z'\}$.
\[
f_{1}=(x)yz+(x)y'z+(1)yz'+(1)y'z'
\]
$f_{2}=x$, $\Phi_{2}(x)=\{x,x'\}$.
Coefficient of $x'$ is zero. Hence consistent. A solution is $x'=1$. That leaves co-efficient of $yz$ equal to zero. Hence assignment $yz=1$ is admissible. This gives $x=0,y=1,z=1$ as one solution.
}
\end{example}

Full details of algorithms based on the above elimination procedure and results of this paper shall be treated in another article. This paper shows an extension of the elimination process following generalization of the one variable Boole-Shannon expansion in terms of an arbitrary ON set $\Phi$. For the special class of functions $B(\Phi)$ when $\Phi$ has polynomial size, it follows that the consistency of $f=0$ can be computed in polynomial number of $B$ operations. The elimination procedure in terms of the general expansion has possible applications for construction of scalable algorithms for computing solutions of Boolean equations especially in the satisfiability problems over the algebra $\bzero$. For brevity of exposition we omit developments of algorithms for special cases of Boolean systems arising in SAT problems for CNFs and DNFs. 

\subsection*{}
\begin{center}
Acknowledgements
\end{center}
Supported by the project grant 11IRCCSG010 of IRCC, IIT Bombay. Author is grateful to Professor Rudeanu for enlightening email correspondence which helped formulation of ideas explored in this paper.

\end{document}